\documentclass[preprint,aps ,nofootinbib]{revtex4}
\usepackage{graphicx}
\pdfoutput=1
\usepackage{epsfig}
\usepackage{amsmath}
\usepackage{amsfonts}
\usepackage{amssymb}
\usepackage{color}%
\usepackage{dcolumn}
\providecommand{\U}[1]{\protect\rule{.1in}{.1in}}

\newcommand{\f}{\begin{equation}}
\newcommand{\ff}{\end{equation}}
\newcommand{\fa}{\begin{eqnarray}}
\newcommand{\ffa}{\end{eqnarray}}
\begin{document}
\title{Holographic Lattice in Einstein-Maxwell-Dilaton Gravity}
\author{Yi Ling $^{1,2}$}
\email{lingy@ihep.ac.cn}
\author{Chao Niu $^{1}$}
\email{niuc@ihep.ac.cn}
\author{Jian-Pin Wu $^{3,4,5}$} \email{jianpinwu@gmail.com}
\author{Zhuo-Yu Xian $^{1}$}
\email{xianzy@ihep.ac.cn}
\affiliation{ $^1$ Institute of High Energy Physics, Chinese
Academy of Sciences, Beijing 100049, China\\
$^2$  State Key Laboratory of Theoretical Physics, Institute of
Theoretical Physics, Chinese Academy of Sciences, Beijing 100190,
China\\
$^3$Department of Physics, School of Mathematics and Physics,
Bohai University,
Jinzhou 121013, China\\
$^4$Department of Physics, Hanyang University, Seoul 133-791, Korea\\
$^5$Center for Quantum Spacetime, Sogang University, Seoul 121-742, Korea}

\begin{abstract}
We construct an ionic lattice background in the framework of
Einstein-Maxwell-dilaton theory in four dimensional space time.
The optical conductivity of the dual field theory on the boundary
is investigated. Due to the lattice effects, we find the imaginary
part of the conductivity is manifestly suppressed in the zero
frequency limit, while the DC conductivity approaches a finite
value such that the previous delta function reflecting the
translation symmetry is absent. Such a behavior can be exactly fit
by the Drude law at low frequency. Moreover, we find that the
modulus of the optical conductivity exhibits a power-law behavior
at intermediate frequency regime. Our results provides further
support for the universality of such power-law behavior recently
disclosed in Einstein-Maxwell theory by Horowitz, Santos and Tong.

\end{abstract} \maketitle

\section{Introduction}

In the past few years, there has been much progress in
understanding strongly coupled condensed matter systems using the
gauge/gravity duality (for recent reviews, see for instance
\cite{Hartnoll:2009Re,Herzog:2009Re,McGreevy:2009Re,Sachdev:2010Re,Hartnoll:2011Re}).
Many interesting phenomena including the superconductivity,
superfluidity and non-Fermi liquid behavior have been addressed in
the holographic framework. Nevertheless, most of these works only
consider homogeneous gravitational backgrounds, which possess
spatial translation invariance, and so the momentum is conserved
in these directions. Due to the conservation of the momentum, many
crucial characteristics of practical condensed matter materials at
zero frequency are far from reached along this route. One known
phenomenon is that a delta function is always present at zero
frequency for the real part of the optical conductivity in the
dual field theory, reflecting an infinite DC conductivity which
obviously is a undesired result for the normal phase of real
materials.

Therefore, in order to build more realistic condensed matter
system in holographic approach, the spatial translational symmetry
in the bulk must be broken such that the momentum in these
directions becomes dissipative. One idea to implement this is
introducing the effects of lattice or impurities in the
gravitational background
\cite{Horowitz:2012ky,Horowitz:2012gs,Horowitz:2013jaa,Zaanen:2012Fe,YiLing:2013Fe,
Hartnoll:2012PRL,Hartnoll:2012Mi,Maeda:2012To,CSPark:2010Mo,Karch:2009La,Karch:2009Di,Sachdev:2013Cr,Mozaffar:2013Cr,Wong:2013Cr,Mozaffar:2013bva,Chesler:2013qla}
\footnote{An alternative approach is considering holographic
massive gravity in which the conservation of momentum is broken by
a graviton mass term\cite{Vegh:2013Ma,Davison:2013Ma,Tong:2013Ma}.
}. From the viewpoint of AdS/CFT correspondence, such a symmetry
breaking in the bulk geometry can lead to significant changes for
some observables in dual field theory on the boundary, for
instance, the optical conductivity and the Fermi surface
structure. Especially, in recent work
\cite{Horowitz:2012ky,Horowitz:2012gs}, a holographic lattice
model has been constructed in the Einstein-Maxwell theory by
introducing a periodic scalar field to simulate the lattice, or
directly imposing a periodic condition for the chemical potential
on the boundary, which is also called as ionic lattice. Several
surprising results have been disclosed by fully solving the
coupled perturbation equations numerically. Firstly, the delta
function at zero frequency spreads out and the conductivity can be
well described by the standard Drude formula at low frequency.
Secondly, in an intermediate frequency regime, a power law
behavior is found as
\begin{eqnarray}
\label{PowerLaw}
|\sigma(\omega)|=\frac{B}{\omega^{\gamma}}+C,
\end{eqnarray}
with $\gamma\simeq 2/3$ in four dimensional spacetime, which is
robust in the sense that it is independent of the parameters of
the model, such as the lattice spacing, lattice amplitude and
temperature. Remarkably, this power law behavior is in good
agreement with those data measured in experiments on the
cuprate\footnote{We note that there is an off-set constant $C$ in
the power law behavior, which is different from that observed in
experiments.} \cite{Marel:2003Na,Marel:2006An}. Another surprising
result is that in the holographic superconducting phase
\cite{Horowitz:2013jaa}, a delta function at zero frequency in the
real part of the optical conductivity does not disappear due to
the presence of the lattice, reflecting that a genuine
superconducting phase is obtained indeed. In particular, the power
law behavior with $|\sigma(\omega)|\propto \omega^{-2/3}$
maintains in the mid-infrared regime of frequency. In addition,
the power law behavior (\ref{PowerLaw}) has also been observed in
holographic massive gravity albeit the exponent $\gamma$ depends
on the mass of the graviton\cite{Vegh:2013Ma,Davison:2013Ma}.

Until now, such power law behavior of (\ref{PowerLaw}) has not
been well understood from the dual gravitational side yet. In this
paper we intend to testify above observations in a different
framework, namely the Einstein-Maxwell-dilaton theory. In this
setup Gubser and Rocha proposed a charged black brane solution
which contains some appealing features for the study of AdS/CMT
correspondence\cite{Gubser:2009qt,Cvetic:1999M5,Gubser:2001M5}.
First of all, such black branes have vanishing entropy in the
extremal case with zero temperature, which is very desirable from
the side of condensed matter materials. Secondly, a linear
dependence of the entropy and heat capacity on temperature at low
temperature can be deduced \cite{Gubser:2009qt,Gubser:1999M5},
which seemingly indicate that the dual field theory is
Fermi-like\cite{JPWu:2011Fe,Gubser:2012Fe,WJLi:2011Fe,XMKuang:2012Fe}
\footnote{A large class of Einstein-Maxwell-dilaton theories with
different Maxwell coupling and scalar potential has been
constructed and they also possess the same appealing
characteristic such as vanishing extremal entropy and linear
specific heat at low temperature (see for instance
\cite{Kiritsis:2010Di,Kachru:2010Di,Pani:2009Di,Iizuka:2011Di}).
But, such solutions are usually not an analytic form.}. The
purpose of our this paper is constructing a lattice background
based on this black brane solution and explore whether such
lattice effects will lead to the same universal behavior of the
optical conductivity as those in RN-AdS black brane
\cite{Horowitz:2012ky,Horowitz:2012gs,Horowitz:2013jaa}.

As the first step, we will concentrate on the lattice effects on
the optical conductivity in normal phase through this paper, but
leave the superconducting phase for further investigation. We will
firstly present a charged dilatonic black brane solution in
Einstein-Maxwell-Dilaton theory in Section \ref{SNoLattice} and
then investigate the optical conductivity without lattice for
later comparison. In Section \ref{SLattice} we explicitly
construct an ionic lattice background by imposing a periodic
chemical potential on the boundary and solving the coupled partial
differential equations numerically with the DeTurck method. Then
we concentrate on the perturbations over this lattice background
and compute the optical conductivity in a holographic approach.
Our conclusions and discussions are given in Section
\ref{SConclusions}.

\section{Conductivity in the charged dilatonic black brane without lattice}\label{SNoLattice}

For comparison, in this section we will calculate the optical
conductivity in the charged dilatonic black brane without lattice.
For relevant discussion on the holographic conductivity in other
backgrounds in Einstein-Maxwell-Dilaton theory, we can refer to,
for instance, \cite{Kiritsis:2010Di,Kachru:2010Di,Pani:2009Di,Salvio:2012at,Salvio:2013jia}.

\subsection{The charged dilatonic black brane}

We begin with the following action, which includes gravity, a
$U(1)$ gauge field, and a dilaton $\Phi$ with a Maxwell non-minimal coupling $f(\Phi)=e^{\Phi}$
and a scalar potential $V(\Phi)=\cosh \Phi$ \cite{Gubser:2009qt}
\f\label{DilatonAction}
S= {1 \over 2\kappa^2}\int d^{4}x \sqrt{-g} \left[ R - {1 \over 4} e^{\Phi} F^{\mu\nu}F_{\mu\nu} -
{3\over 2} (\partial_\mu\Phi)^2 + {6 \over L^2} \cosh \Phi \right].
\ff
Applying the variational approach to the above action, one can
obtain the equations of motion for the fields $g_{\mu\nu}$,
$A_{\mu}$ and $\Phi$, respectively
\fa \label{EOMg}
&&
R_{\mu\nu}
=\frac{1}{2}e^{\Phi}(F_{\mu\sigma}{F_\nu}^\sigma-\frac{1}{4}F^2g_{\mu\nu})+
\frac{3}{2}\partial_{\mu}\Phi\partial_{\nu}\Phi-\frac{3}{L^2}g_{\mu\nu}\cosh \Phi,
\\ \label{EOMA}
&&
\nabla_{\mu}\left(e^{\Phi}F^{\mu\nu}\right)=0,
\\ \label{EOMPhi}
&&
\Box^{2}\Phi=\frac{1}{12}e^{\Phi}F^2-\frac{2}{L^2}\sinh \Phi,
\ffa
where $\Box^{2}\equiv
\nabla^{\mu}\nabla_{\nu}=g^{\mu\nu}\nabla_{\mu}\nabla_{\nu}$ is
the covariant d'Alembertian operator. An analytical charged
solution to these equations has previously been given in
\cite{Gubser:2009qt}, which reads as
\fa \label{Metric}
&&
ds^{2} = \frac{L^2}{z^2}\left(-f(z)dt^2+\frac{dz^2}{f(z)}+g(z)(dx^2+dy^2)\right),
\\ \label{GaugeAt}
&& A_{t}(z)=L\sqrt{3Q}(1-z)\frac{\sqrt{1+Q}}{1+Q z},
\\ \label{Phi}
&&
\Phi(z)=\frac{1}{2}\ln\left(1+Q z\right),
\ffa
where
\fa
&&
f(z)=(1-z)\frac{p(z)}{g(z)},~~~~g(z)=\left(1+Qz\right)^{3/2},
~~~~
\nonumber\\
&&\label{Metricp}
p(z)=1+\left(1+3Q\right)z+\left(1+3Q\left(1+Q\right)\right)z^{2}.
\ffa
Here for the convenience of the numerical analysis, we have made a
coordinate transformation based on the solution presented in
\cite{Gubser:2009qt}\footnote{Here after having fixed the radius
of the horizon at $z=1$, we have only one parameter $Q$, which as
a matter of fact corresponds to $q/r_h$ in \cite{Gubser:2009qt},
where $q$ is the charge and $r_h=(m L^2)^{1/3}-q$ is the position
of the horizon with mass $m$. In \cite{Gubser:2009qt} the extremal
limit is given by $q=(m L^2)^{1/3}$, which in our case corresponds
to $Q\rightarrow \infty$ since $ r_h\rightarrow 0$. }. Next we are
concerned with the Hawking temperature $T$ of this charged black
brane. A simple algebra shows that in our current coordinate
system, the Hawking temperature is \f \label{T} T
=\frac{3\sqrt{1+Q}}{4\pi L }. \ff

However, due to the conformal symmetry on the boundary, we stress
that whatever the coordinates system is adopted, the quantity
which is physically meaningful is the ratio of $T/\mu$ in
asymptotically charged AdS space time, namely the temperature
measured with the chemical potential as the unit. This quantity is
invariant under the rescaling of the time coordinate. Therefore,
through this paper we will use the chemical potential on the
boundary to set the unit of the system. In our case we find the
Hawking temperature measured with the chemical potential reads as
\f \label{HawkingT} \frac{T}{\mu} =\frac{\sqrt{3}}{4\pi L
\sqrt{Q}}. \ff
Obviously, this temperature with the chemical potential as the
unit is a monotonically decreasing function of $Q$. As $Q$ tends
to zero, it approaches to infinity, which corresponds to a
Schwarzschild-AdS black brane. On the other hand, as $Q$ goes up
to infinity, the ratio $T/\mu$ runs to zero, corresponding to an
extremal black hole. An appealing characteristic of this black
brane solution is that its entropy on the horizon is vanishing in
the extremal case, in contrast to the usual RN-AdS black holes
which approach a $AdS2$ geometry near the horizon. As a result,
this sort of black brane is expected to be a more practical arena
to implement a holographic scenario for a condensed matter system
at low temperature. For more discussion on the thermodynamics of
this charged dilatonic black brane, please refer to
\cite{Gubser:2009qt}. In this end of this subsection, we remark
that our current coordinate system may not be appropriate for
numerical analysis in zero temperature limit since in this case
the parameter $Q$ goes to infinity. We need transform the metric
into some form closely following the treatment as given in
\cite{Horowitz:2013jaa}. Nevertheless, since in this paper we are
not concerned with its zero temperature behavior, we argue that
this coordinate system is good enough for us to observe the
lattice effects in a rather large regime of the temperature.

\subsection{Conductivity}

To compute the conductivity in a holographic approach, the
simplest way is turning on the gauge field fluctuation $a_{x}$ and
the metric fluctuation $h_{tx}$. Before proceeding, we take a
Fourier transformation for the perturbed fields
\begin{eqnarray}
\label{gaFourier} h_{tx}(t,z)\sim e^{-i\omega
t}h_{tx}(z),~~~~~a_{x}(t,z)\sim e^{-i\omega t}a_{x}(z).
\end{eqnarray}
Here, for simplicity, we stay at zero momentum $k=0$. Perturbing
the equations of motion in (\ref{EOMg}) and (\ref{EOMA}) at the
linear level gives the following two independent equations
\begin{eqnarray}
\label{EM3}
&&
a''_{x}+\left(\frac{f'}{f}+\Phi'\right)a'_{x}+\frac{\omega^{2}}{f^{2}}a_{x}
+\frac{z^{2}A'_{t}}{f}h'_{tx}+\left(\frac{2zA'_{t}}{f}-\frac{z^{2}A'_{t}g'}{f g}\right)h_{tx}=0,
\\ \label{GdS23}
&&
h'_{tx}+\left(\frac{2}{z}-\frac{g'}{g}\right)h_{tx}+\frac{e^{\Phi}A'_{t}}{L^2}a_{x}=0.
\end{eqnarray}
Substituting the second equation into the first one, one can
easily obtain a decoupled equation for the fluctuations of the
gauge field $a_{x}$ which is
\begin{eqnarray}
\label{ax}
a''_{x}+\left(\frac{f'}{f}+\frac{\Phi'}{\sqrt{3}}\right)a'_{x}+\left(\frac{\omega^{2}}{f^{2}}-\frac{z^{2}e^{\Phi}{A'_{t}}^2}{L^{2}f}\right)a_{x}=0.
\end{eqnarray}
We can solve the above equation with purely ingoing boundary
conditions at the horizon ($z\rightarrow 1$). And then, the
conductivity can be read off from the formula
\begin{eqnarray}
\label{Conductivity}
\sigma(\omega)=-i\frac{a_{x}^{(1)}}{\omega a_{x}^{(0)}},
\end{eqnarray}
where $a^{(0)}_{x}$ and $a^{(1)}_{x}$ are determined by the
asymptotic behavior of the fluctuation $a_{x}$ at the boundary
($z\rightarrow 0$)
\begin{eqnarray}
\label{axz0}
a_{x}=a^{(0)}_{x}+a^{(1)}_{x}z+\ldots.
\end{eqnarray}
%

\begin{figure}
\center{
\includegraphics[scale=0.6]{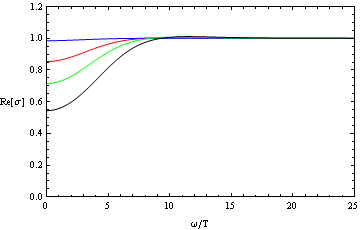}\hspace{0.1cm}
\includegraphics[scale=0.6]{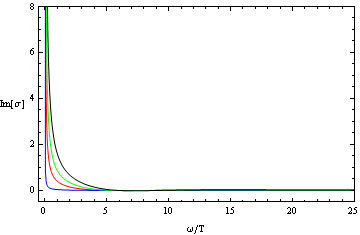}\\ \hspace{0.1cm}
\caption{\label{ReIm}The real (left) and imaginary (right) parts of the conductivity for different $Q$
(blue for $Q=0.01$, red for $Q=0.1$, green for $Q=0.25$ and black for $Q=0.5$).
}}
\end{figure}

The numerical results for the real and imaginary parts of the
conductivity are shown in FIG.\ref{ReIm}. In the high-frequency
limit, the real part of the conductivity becomes constant. It is a
common characteristic for the holographic dual theory. However,
the imaginary part of the conductivity goes to infinity at
$\omega=0$ and the standard Kramers-Kronig relation implies that
there is a delta function at $\omega=0$ in the real part of the
conductivity, which is of course not a genuine superconductivity
phase but a consequence of translational symmetry. At small
frequencies, some interesting behaviors exhibit in the real part
of the conductivity. From FIG.\ref{ReIm}, we can observe that it
develops a minimum and the minimum increases with the decrease of
$Q$. Finally, it almost becomes a constant when $Q\rightarrow 0$,
which is just the case for the Schwarzschild-AdS black brane.

\section{Conductivity in the charged dilatonic black brane with lattice}\label{SLattice}

In this section, we will firstly construct an ionic lattice
background in Einstein-Maxwell-Dilaton theory, and then
demonstrate how the optical conductivity would be greatly changed
in low frequency limit due to the presence of the lattice.

\subsection {A Holographic Charged Dilaton Lattice}

As the first step, through this paper we will introduce the
lattice only in $x$ direction but leave the $y$ direction
translation invariance. As a result, one may take the following
ansatz for the background metric, which is compatible with the
symmetry we considered above
\f
ds^2=\frac{L^2}{z^2}\left[-(1-z)p(z)H_1dt^2+\frac{H_2dz^2}{(1-z)p(z)}+S_1
(dx+Fdz)^2+S_2 dy^2\right], \label{eq:line1} \ff with \f A_t
=L\sqrt{3Q}(1-z)\psi(x,z), \ff
and
\f\label{eq:line3} \Phi
=\frac{1}{2}\ln\left(1+Q z\phi(x,z)\right), \ff
where $H_{1,2},S_{1,2},F,\psi $ and $\phi$ are seven functions of
$x$ and $z$, and will be determined by solving Eqs.(\ref{EOMg}),
(\ref{EOMA}) and (\ref{EOMPhi}). Note that if
$H_1=1/g(z),~H_2=S_1=S_2=g(z),~F=0,~\psi=\sqrt{1+Q}/(1+Q z)$ and
$\phi=1$, it will recover to the charged dilatonic black brane
solution without lattice described by
(\ref{Metric}-\ref{Metricp}).

Now, we introduce the lattice by imposing an inhomogeneous
boundary condition for the chemical potential as
\f
\psi(x,0)=\sqrt{1+Q}(1+A_0\cos(k_0x)),
\ff
which is often referred to as an ionic lattice.

\begin{figure}
\center{
\includegraphics[scale=0.275]{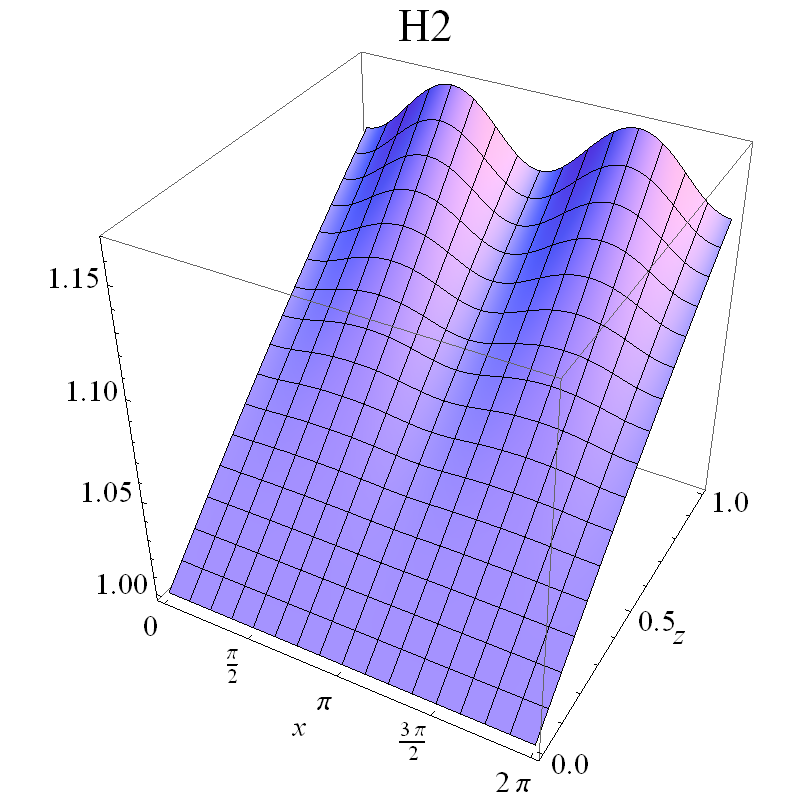}\hspace{0.1cm}
\includegraphics[scale=0.275]{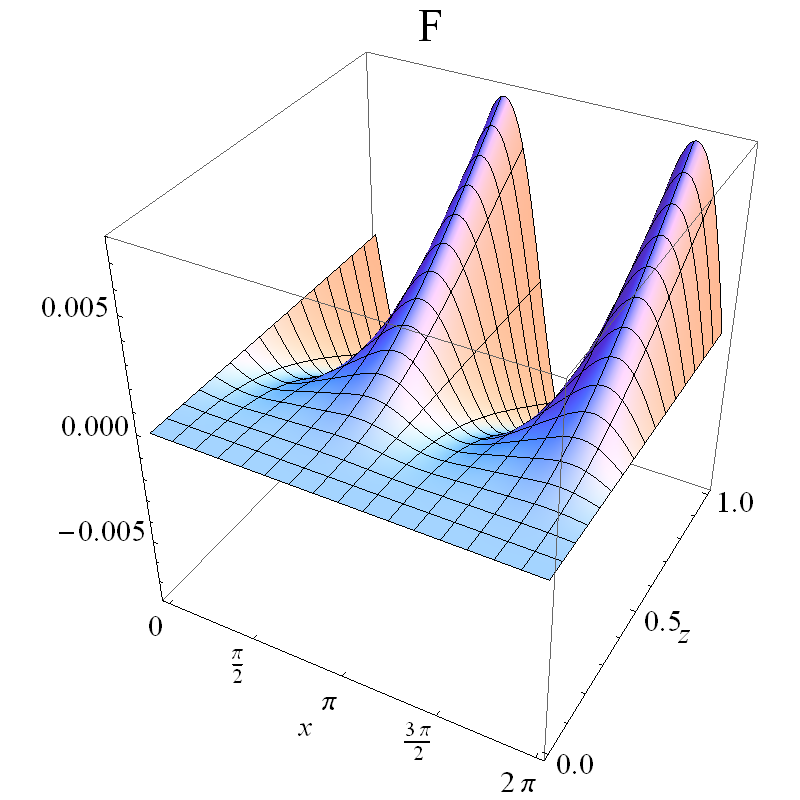}\\ \hspace{0.1cm}}
\center{
\includegraphics[scale=0.275]{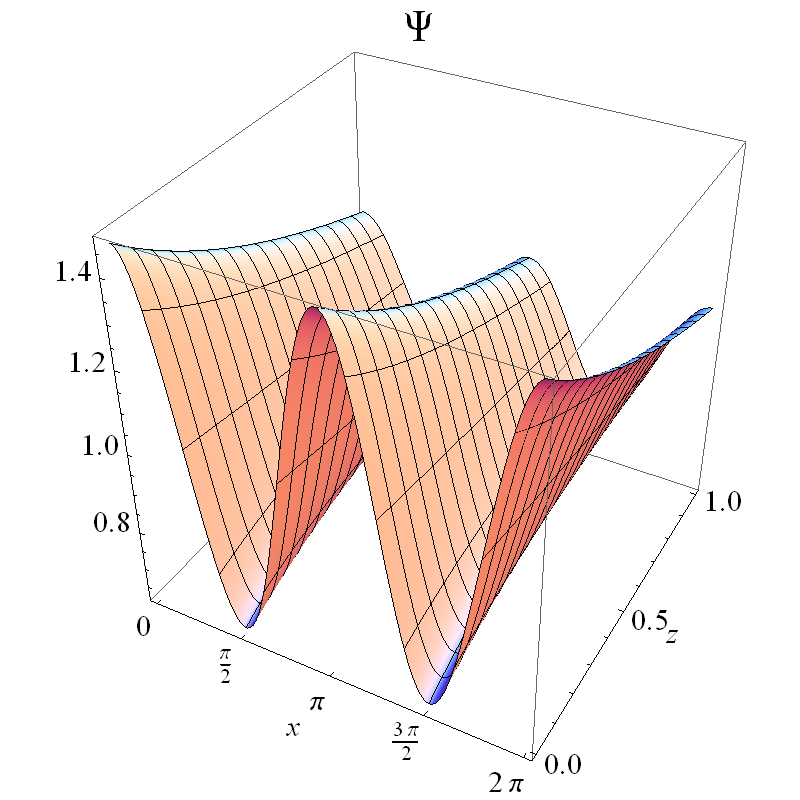}\hspace{0.1cm}
\includegraphics[scale=0.275]{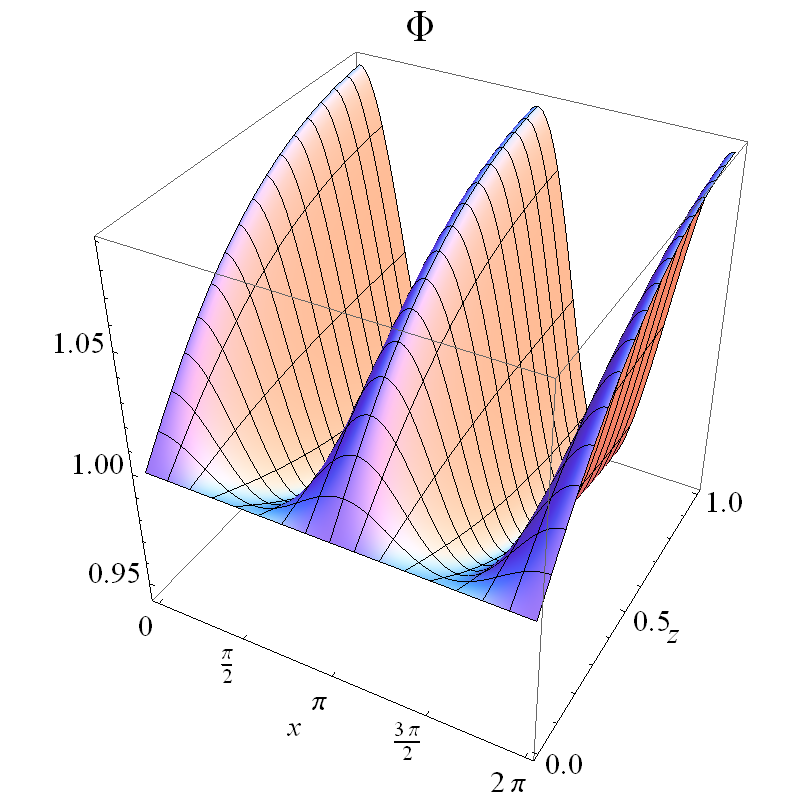}\\ \hspace{0.1cm}
\caption{\label{fig:background}We show $H_2$, $F$, $\psi$ and
$\phi$ for $k_0=2$, $A_0 =0.4$, $Q = 0.1$ and $|T/\mu| = 0.457$.}
}
\end{figure}

The nonlinear PDEs (\ref{EOMg}-\ref{EOMPhi}), together with the
ansatz (\ref{eq:line1}-\ref{eq:line3}), are solved using the
Einstein-DeTurck method\cite{Headrick:2009pv}. We choose the
charged dilatonic black brane (\ref{Metric}-\ref{Metricp}) as the
reference metric. We firstly linearize the PDEs
(\ref{EOMg}-\ref{EOMPhi}) with the use of Newton-Kantorovich
method and then change the differential equations into algebraic
equations employing the spectral collocation method. Such a
boundary value problem will be approximated by Fourier
discretization in $x$ direction and Chebyshev polynomials in $z$
direction. As an example, we show a solution in
FIG.~\ref{fig:background} with $A_0=0.4$, $k_0=2$ and $Q=0.1$,
which corresponds to a lattice with $T/\mu=0.457$.

\begin{figure}
\center{
\includegraphics[scale=0.28]{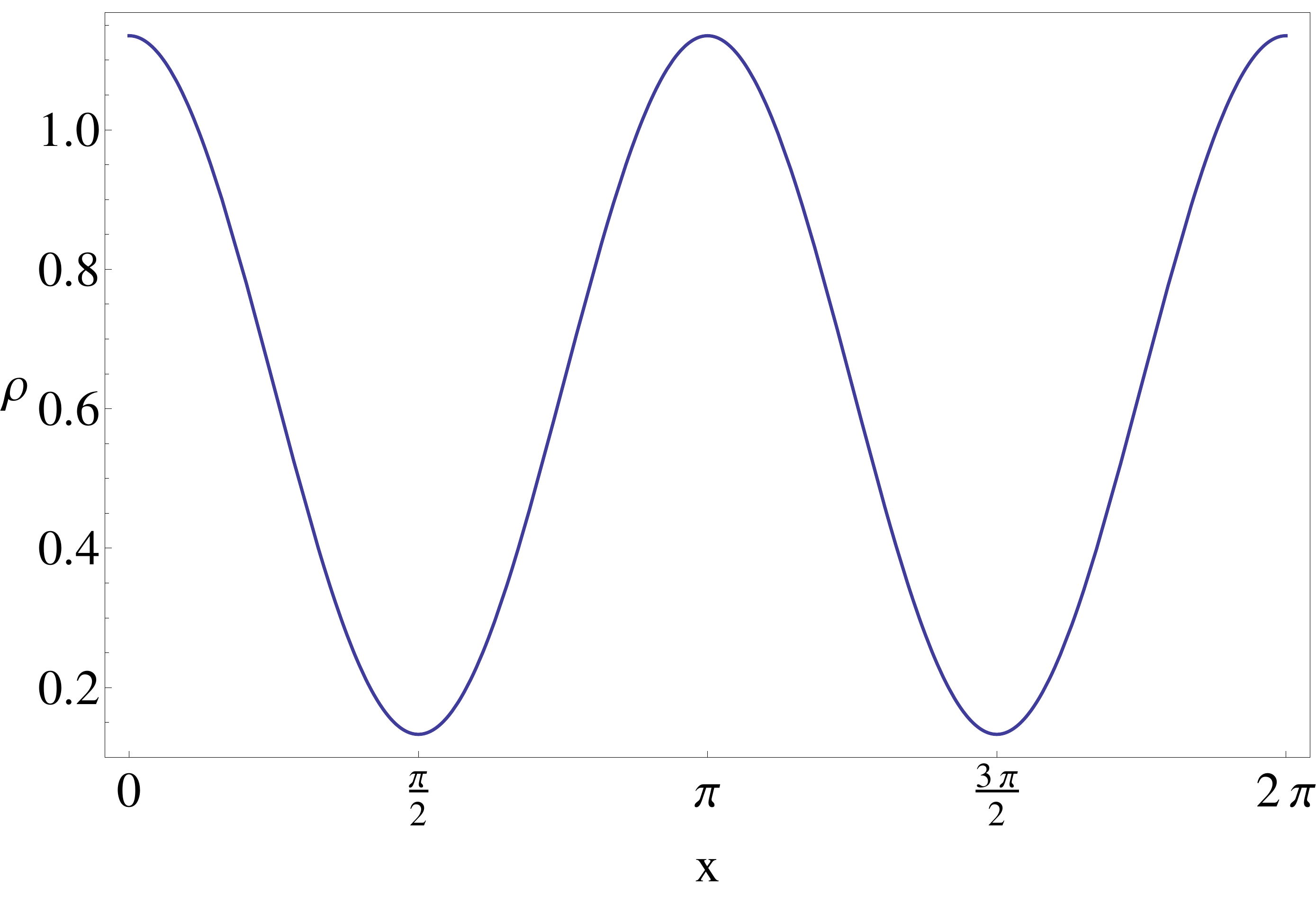}\hspace{0.1cm}
\includegraphics[scale=0.28]{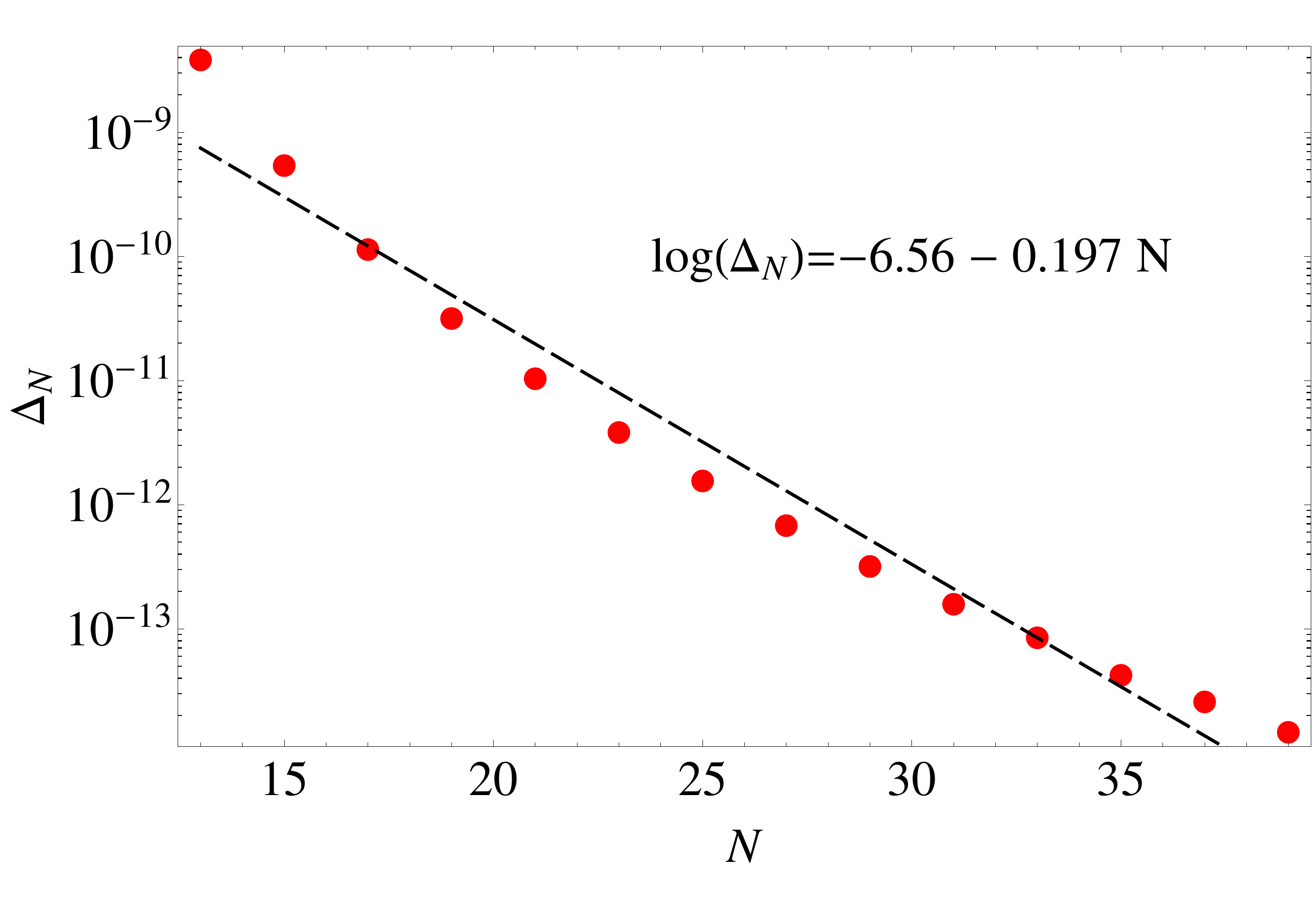}\\ \hspace{0.1cm}
\caption{\label{tendency}On the left side we show the charge
density $\rho$ as a function of $x$ on the boundary, which can be
read off by expanding $\psi=\mu+(\mu-\rho)z+\mathcal{O}(z^2)$. On
the right we show the charge discrepancy $\Delta_N$ as a function
of the number of grid points $N$, where the boundary charge is
evaluated in a period as $Q=\int_0^{2\pi/k_0}dx \rho$. The
vertical scale is logarithmic, and the data is well fit by an
exponential decay: $\log(\Delta_N) = -6.56-0.197\,N$.} }
\end{figure}

In order to guarantee the convergence of our numerical solutions,
we have demonstrated the decaying tendency of the charge
discrepancy $\Delta_N$ which is defined as
$\Delta_N=|1-Q_N/Q_{N+1}|$, where $Q_N$ is the charge on the
boundary with $N$ grid points. As shown in FIG.~\ref{tendency}, an
exponential decay implies our solutions are exponentially
convergent with the increase of grid points. Moreover, we have
also checked the behavior of the DeTurck vector field $\xi^a$
which is defined in \cite{Horowitz:2012ky} and found that for our
solution its magnitude is controlled below $10^{-10}$ indeed. In
this sense our lattice configuration is just a ripple over the
previous charged black brane.

\subsection{Conductivity}

To explore the lattice effects on the optical conductivity in a
holographic approach, we need consider the perturbations over the
lattice background with full backreactions. Therefore, we need in
principle turn on all the gravitational fields as well as gauge
and dilaton fields, and allow them to fluctuate with a harmonic
frequency. Since we only take the lattice into account in
$x$-direction, one economical way in numerical analysis is
assuming that all perturbation quantities do not depend on the
coordinate $y$. This ansatz reduces the perturbation variables to
$11$ unknown functions $\{h_{tt}, h_{tz}, h_{tx}, h_{zz}, h_{zx},
h_{xx}, h_{yy}, b_t, b_z, b_x, \eta\}$ and all of them are just
dependent of $x$ and $z$. For instance, we may have
\begin{eqnarray}
\delta A_x=b_{x}(t,x,z)\sim e^{-i\omega t}b_{x}(x,z).
\end{eqnarray}
Plugging all the perturbations into the equations of motion and
ignoring the higher order corrections, we obtain 11 linear partial
differential equations. In addition, we adopt the harmonic gauge
conditions and Lorentz gauge condition for gravity and Maxwell
field, respectively. The last thing is to set the suitable
boundary conditions. We find the most economical but self
consistent way to do this at $z=0$ is requiring the x-component of
the Maxwell field to be
\begin{eqnarray}
b_{x}(x,z)=1+j_x(x)z+{\cal O}(z^2),
\end{eqnarray}
while the other perturbations are vanishing at the zeroth order of
$z$. Of course we point out that such a choice is not unique, for
instance we are allowed to set the dilaton perturbation to be a
non-zero constant on the boundary, the corresponding solution can
be found and we find such adjustment will not affect our following
observations on the optical conductivity. Thus, for simplicity we
set $\eta (x,0)=0$. As far as the boundary conditions at $z=1$ are
concerned, we adopt the standard regularity conditions on horizon
and consider only the ingoing modes. In addition, in numerical
analysis the highly oscillating modes near the horizon can be
filtered by changing the variables with a factor such as
$(1-z)^{-i\omega/4\pi T}$, and we refer to \cite{Horowitz:2012ky}
for detailed discussions.

\begin{figure}
\center{
\includegraphics[scale=0.275]{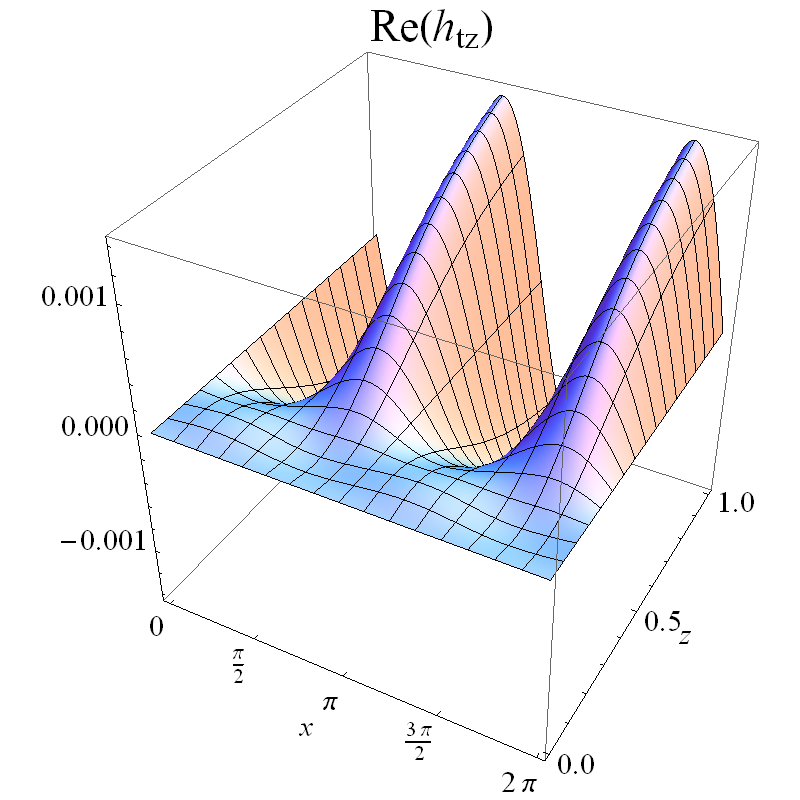}\hspace{0.1cm}
\includegraphics[scale=0.275]{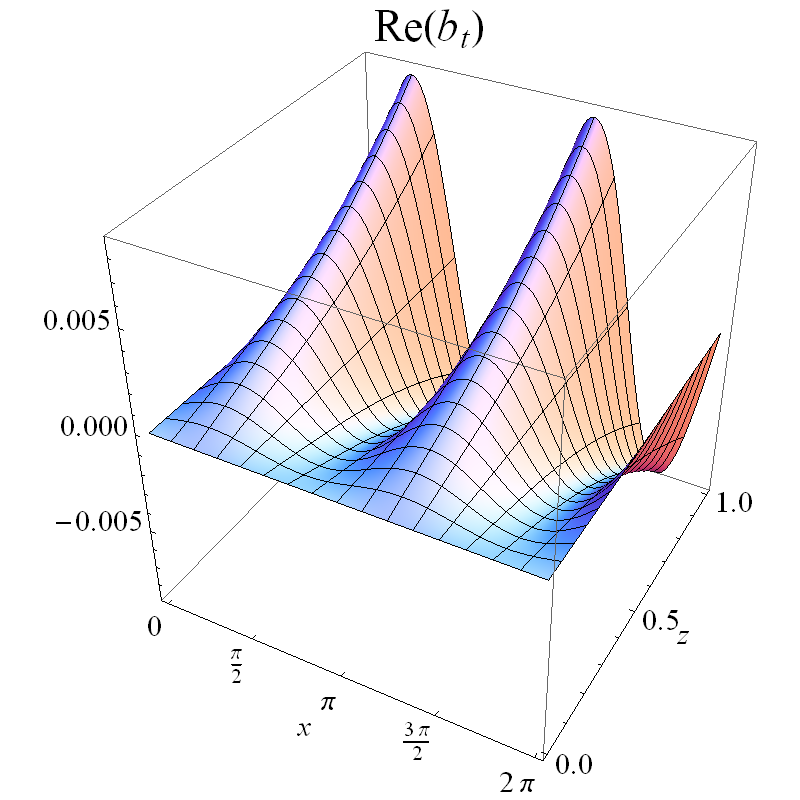}\\ \hspace{0.1cm}}
\center{
\includegraphics[scale=0.275]{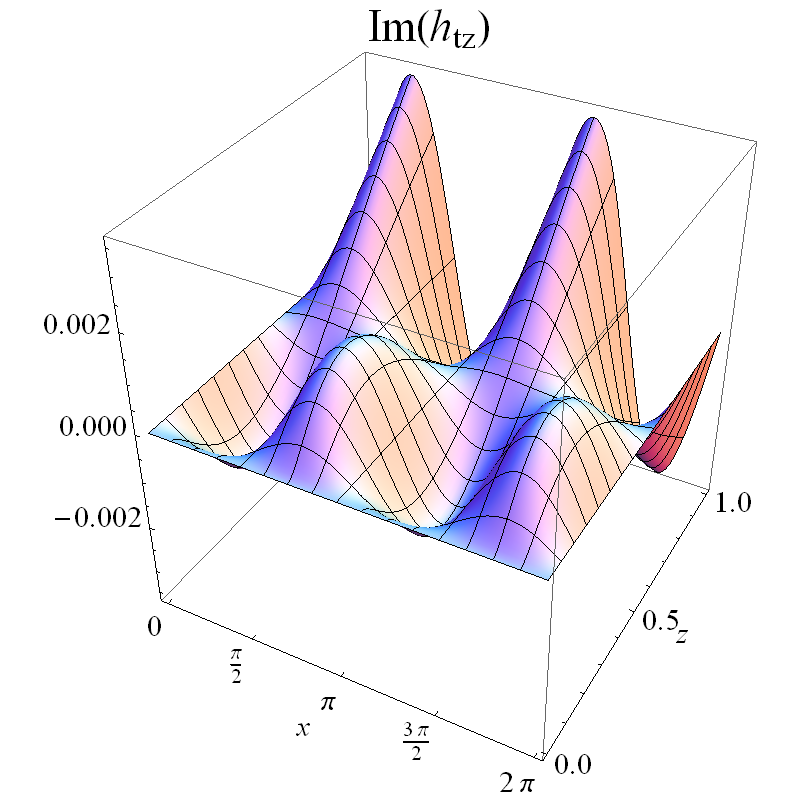}\hspace{0.1cm}
\includegraphics[scale=0.275]{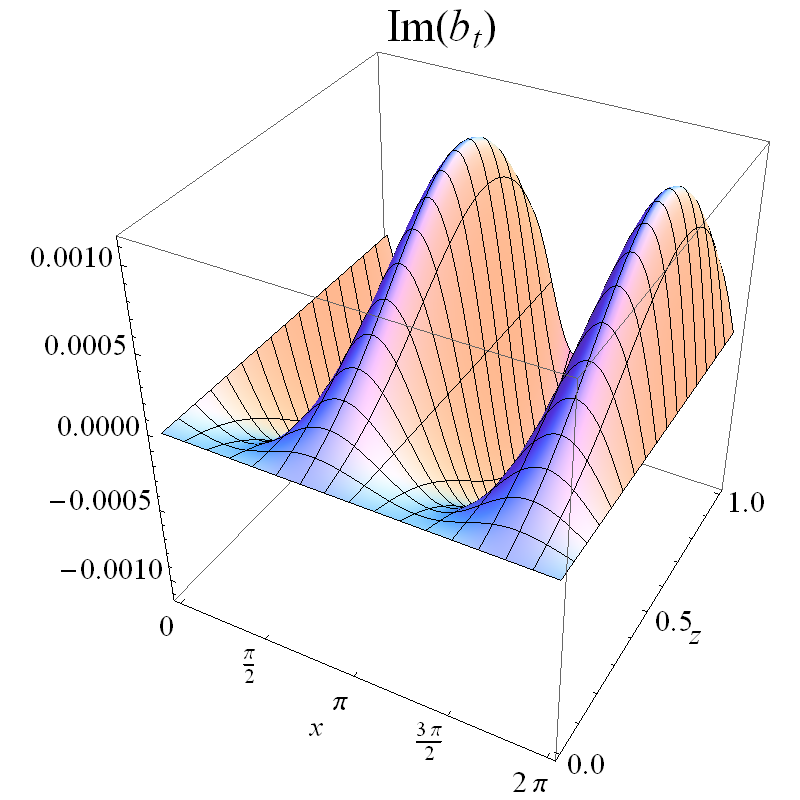}\\ \hspace{0.1cm}
\caption{\label{fig:perturbation}The real and imaginary parts of $h_{tz}$ and $b_x$ are shown for $A_0=0.4$, $k_0=2$, $|T/\mu|=0.457$ and $\omega=0.6$.}
}
\end{figure}

Finally, we solve all these linear equations using a standard
pseudo-spectral collocation methods. In
FIG.~\ref{fig:perturbation}, we show a solution for $h_{tz}$ and
$b_x$ with $A_0=0.4$, $k_0=2$, $T/\mu=0.457$ and $\omega=0.6$.

The optical conductivity in the direction of the lattice can be
read off
\begin{equation}
\sigma(\omega,x)=\frac{j_x(x)}{i \omega}.
\end{equation}
Note that the conductivity is a function of $x$. Since the
boundary electric field $E_x=i \omega e^{-i\omega t}$ is
homogeneous, the homogeneous part of the conductivity is the
quantity we study below.

\begin{figure}
\center{
\includegraphics[scale=0.28]{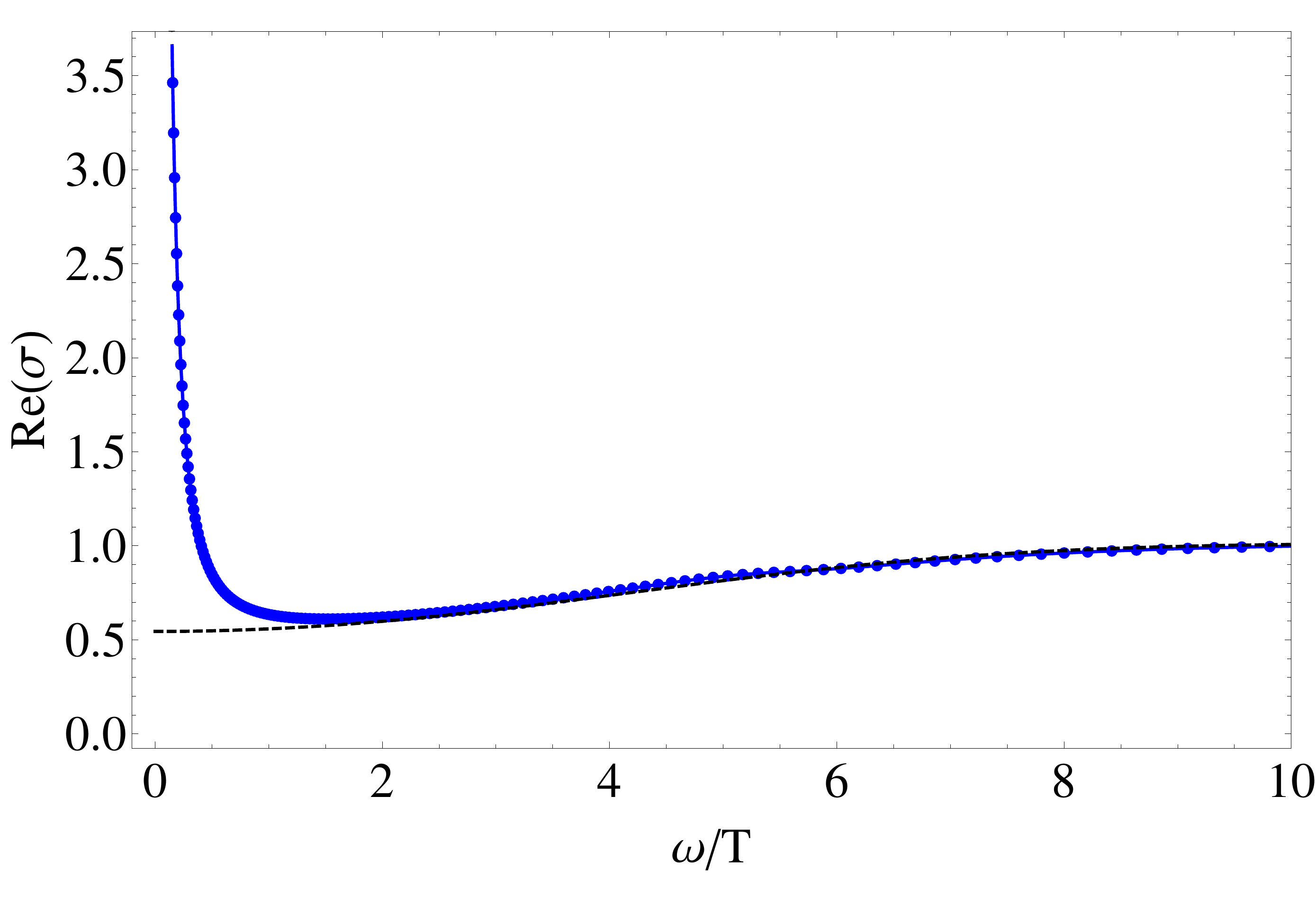}\hspace{0.1cm}
\includegraphics[scale=0.28]{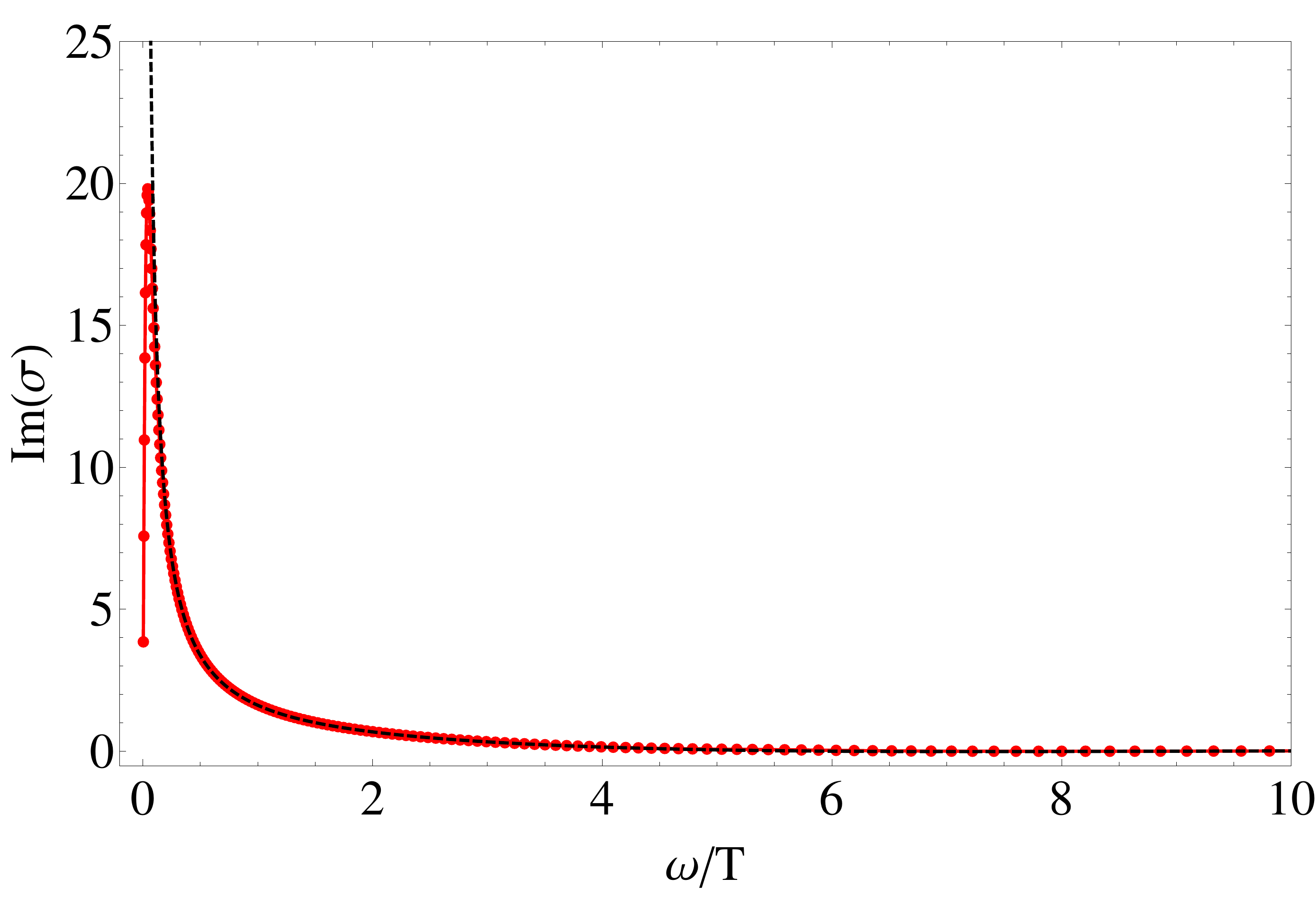}\\ \hspace{0.1cm}
\caption{\label{conductivity}The real and imaginary parts of the conductivity, both without the lattice (dashed line) and with the lattice (solid line and data points) for $A_0=0.4$, $k_0=2$ and $Q=0.5$. Only the low frequency regime of the conductivity is changed by the lattice.}
}
\end{figure}

\begin{figure}
\center{
\includegraphics[scale=0.28]{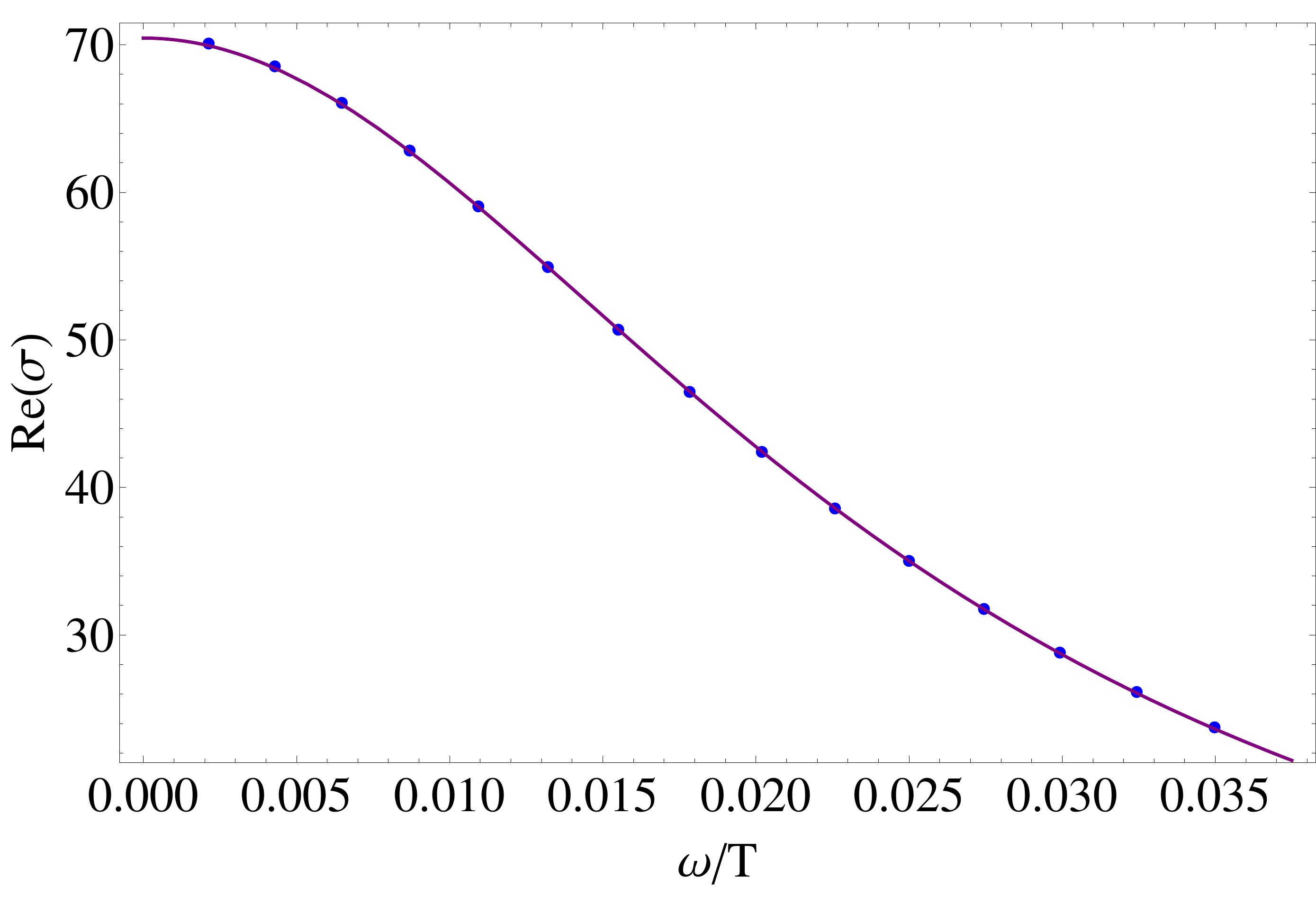}\hspace{0.1cm}
\includegraphics[scale=0.28]{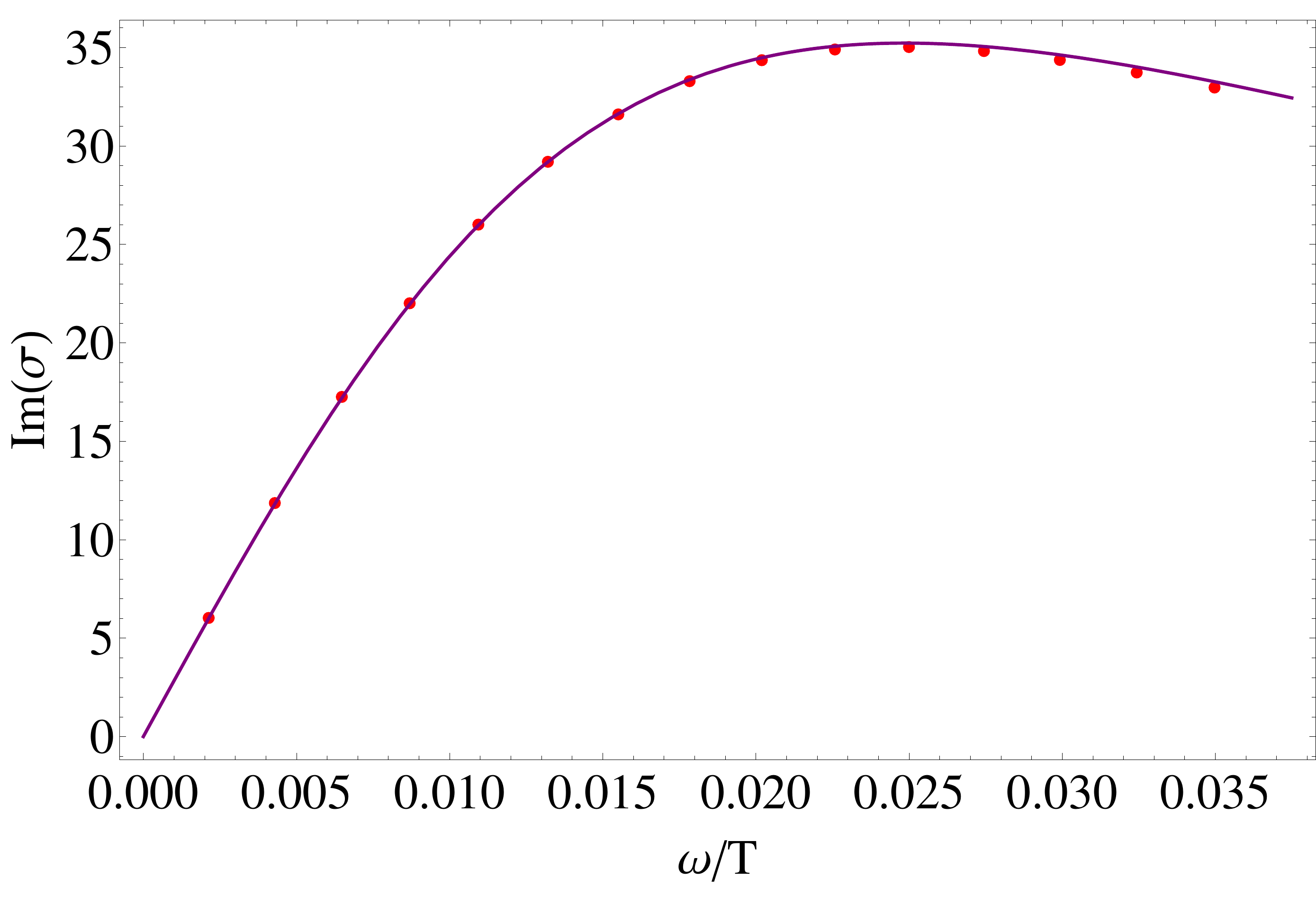}\\ \hspace{0.1cm}
\caption{\label{Drude}A blow up of the low frequency part of the
conductivity with $A_0=0.3$, $k_0=2$ and $Q=0.5$. The data can be
fit by the simple two parameter Drude form.} }
\end{figure}

A typical effect due to the presence of lattice on the optical
conductivity is demonstrated in FIG.~\ref{conductivity}. The
dashed lines represent the conductivity without lattice while the
solid lines represent the conductivity at the same temperature
with lattice.
Obviously, in high frequency regime the conductivity with lattice
coincides with that without lattice. However, in low frequency
regime the conductivity is greatly changed by the lattice. We
observe that the imaginary part of the conductivity has been
significantly suppressed, while the previous delta function at
zero frequency for the real part spreads out, which is consistent
with the Kramers-Kronig relation. Quantitatively we find such
suppression can be well fit by the famous Drude law with two
parameters
\begin{equation}
\sigma(\omega) =\frac{K \tau}{1-i\omega\tau},\label{drude}
\end{equation}
where $\tau$ is the scattering time and $K$ is an overall constant
characterizing the DC conductivity. A blow up of the low frequency
part is shown in FIG.~\ref{Drude} with $A_0=0.4$, $k_0=2$ and
$Q=0.5$. No matter how we change the temperature or lattice
spacing, we find the Drude law can hold in all cases we have
examined.

\begin{figure}
\center{
\includegraphics[scale=0.285]{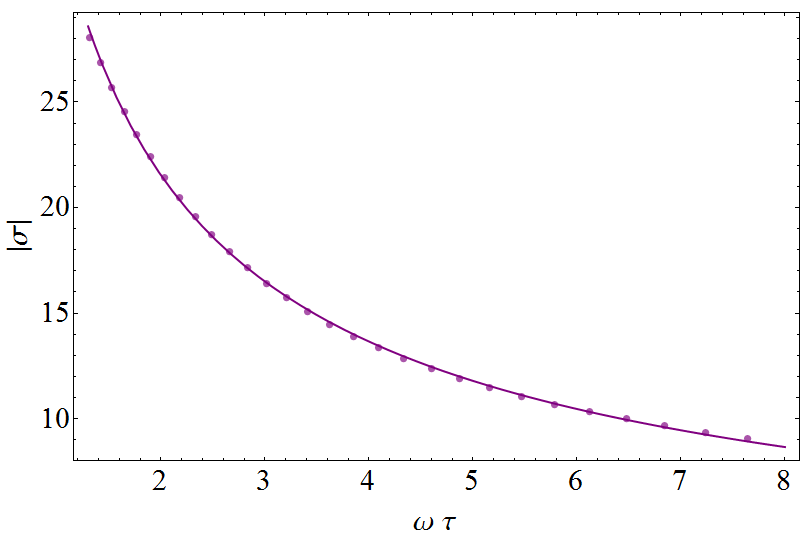}\hspace{0.1cm}
\includegraphics[scale=0.285]{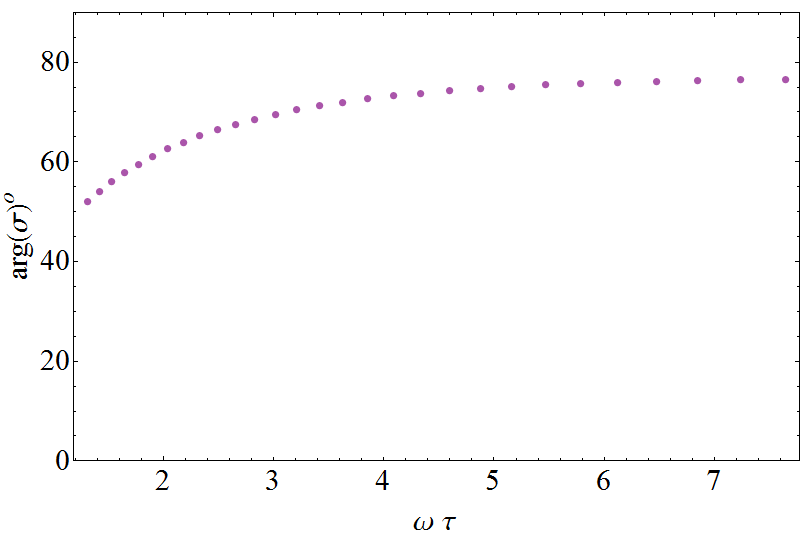}\\ \hspace{0.1cm}
\caption{\label{powerlaw1}The modulus $|\sigma|$ and argument
arg$\sigma$ of the conductivity with $A_0=0.4$, $k_0=2$ and
$Q=0.5$. The line of the left is a fit to the power-law in
(\ref{power}) for $1.3\leq\omega \tau\leq 8$.} }
\end{figure}

\begin{figure}
\center{
\includegraphics[scale=0.37]{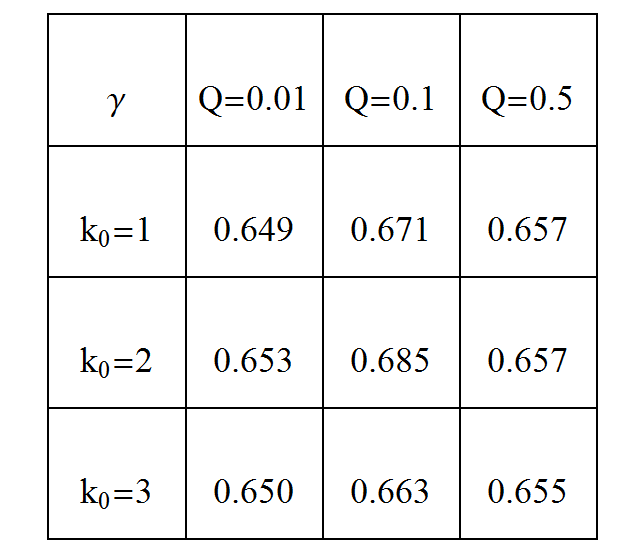}\hspace{0.1cm}
\includegraphics[scale=0.29]{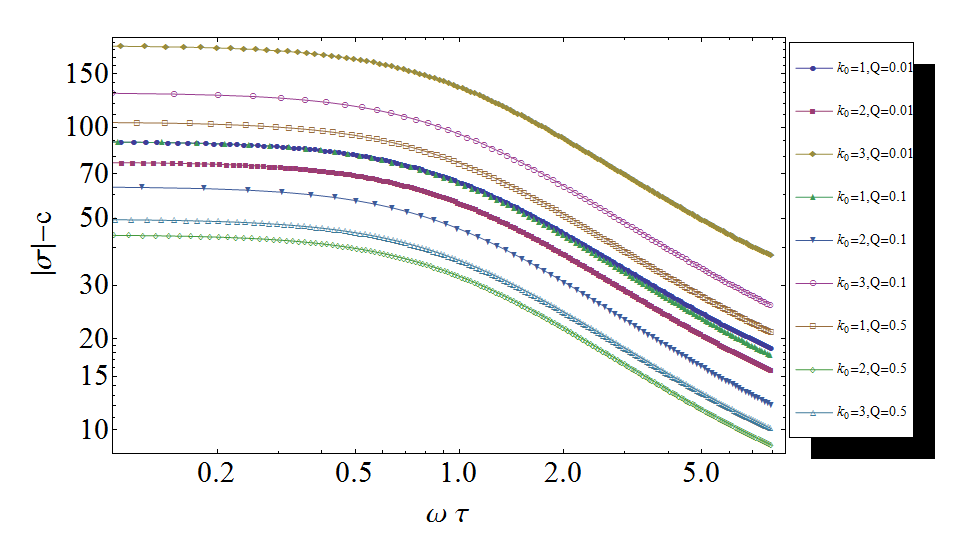}\\ \hspace{0.1cm}
\includegraphics[scale=0.37]{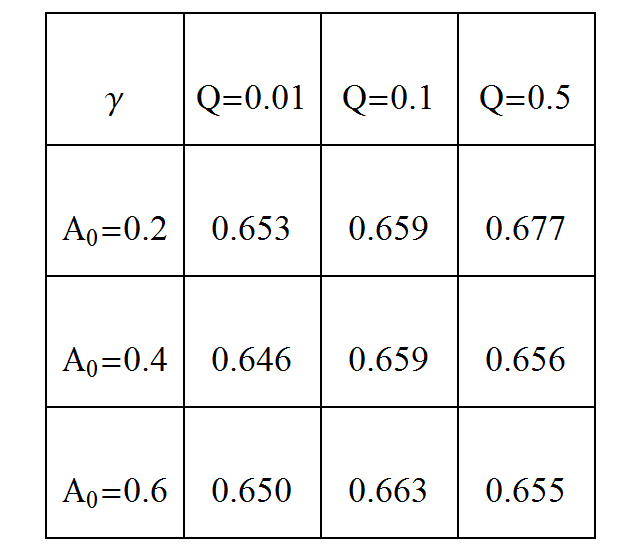}\hspace{0.1cm}
\includegraphics[scale=0.287]{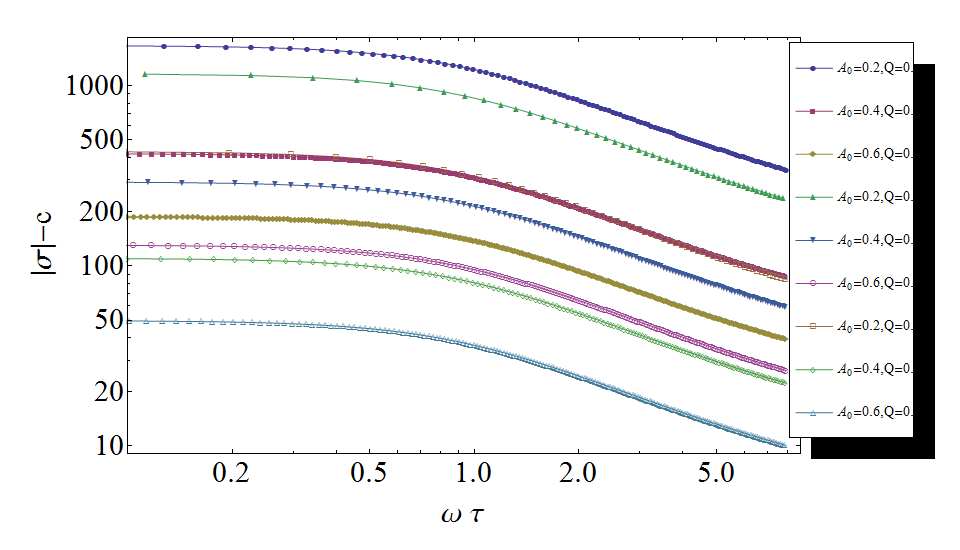}\\ \hspace{0.1cm}
\caption{\label{gamma1}The values of the exponent $\gamma$ with
different $Q$, $A_0$ and $k_0$ for $1.3\leq\omega \tau \leq8$. The
table and the figure on the top have $k_0/A_0=5$, whereas the
table and the figure at the bottom have $k_0=3$.} }
\end{figure}

From Drude formula one easily finds that the peak of the imaginary
part of the conductivity appears at $\omega \tau=1$. When $\omega
\tau>1$, the frequency enters an intermediate regime, in which we
find the modulus of the conductivity exhibits a power-law
fall-off. In FIG.~\ref{powerlaw1} we plot the modulus and argument
of the conductivity in this regime and then fit the data with a
power-law formula
\begin{equation}
|\sigma(\omega)|  = \frac{B}{\omega^{\gamma}} + C,\label{power}
\end{equation}
which contains three free parameters $B$,$C$ and $\gamma$.
Remarkably, when we fix the frequency regime to be $1.3\leq\omega
\tau\leq 8$, we find the exponent $\gamma \simeq 2/3$ is a
universal result, irrespective of what temperature, lattice
amplitude or spacing we have taken. We collect our fitting data
for the exponent in a table for various values of parameters,
accompanying with a log plotting of the variation of the
conductivity with the frequency in FIG.~\ref{gamma1}, in which it should be
noticed that for each case an offset $C$ with different value has
been subtracted from the vertical axis. This universality further
supports the results originally presented in
\cite{Horowitz:2012gs}. Our results indicate that such a
universality may be extended to various lattice models in four
dimensional spacetime, implying the exponent might be singly
related to the dimensionality of the space time.

\begin{figure}
\center{
\includegraphics[scale=0.28]{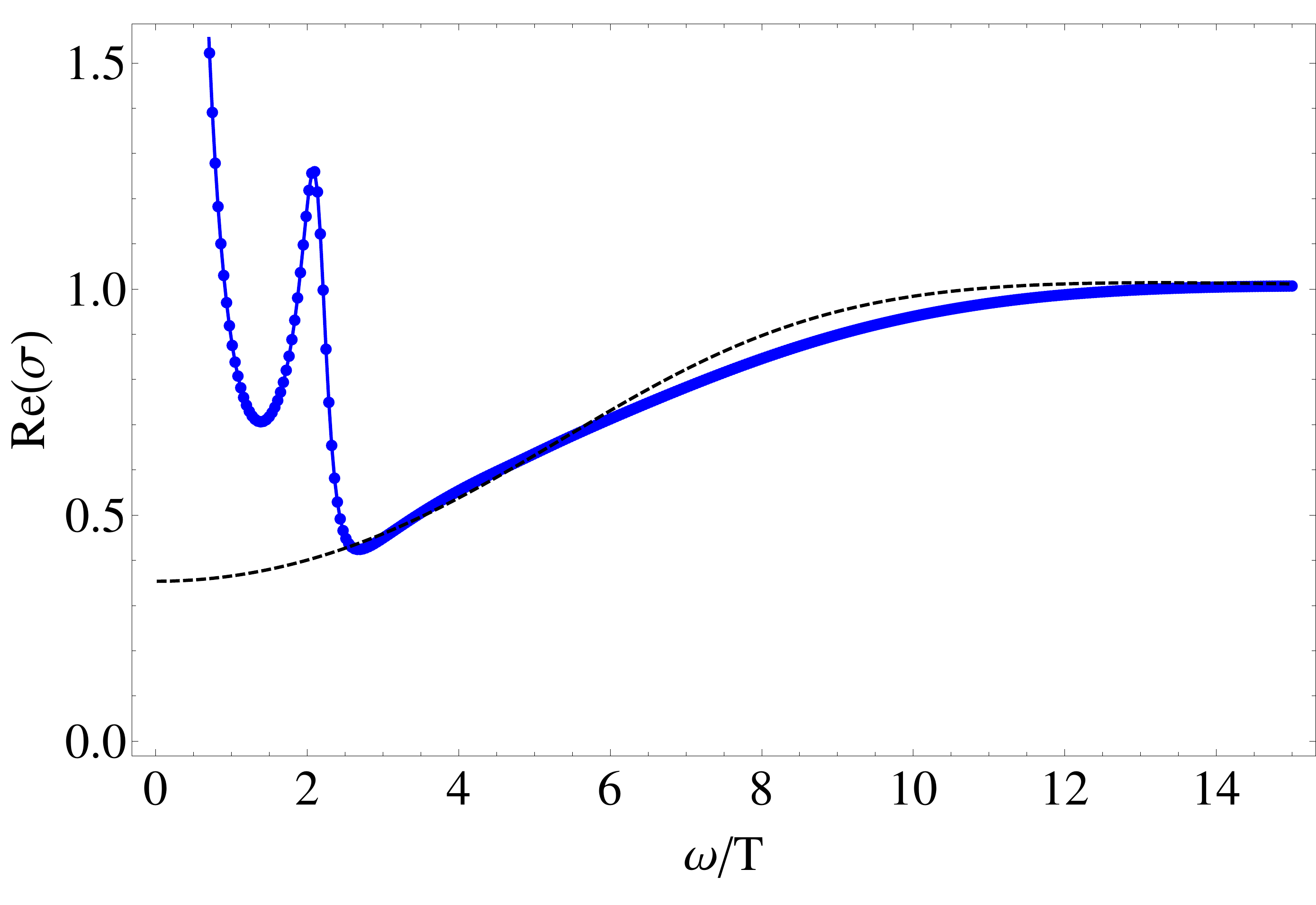}\hspace{0.1cm}
\includegraphics[scale=0.28]{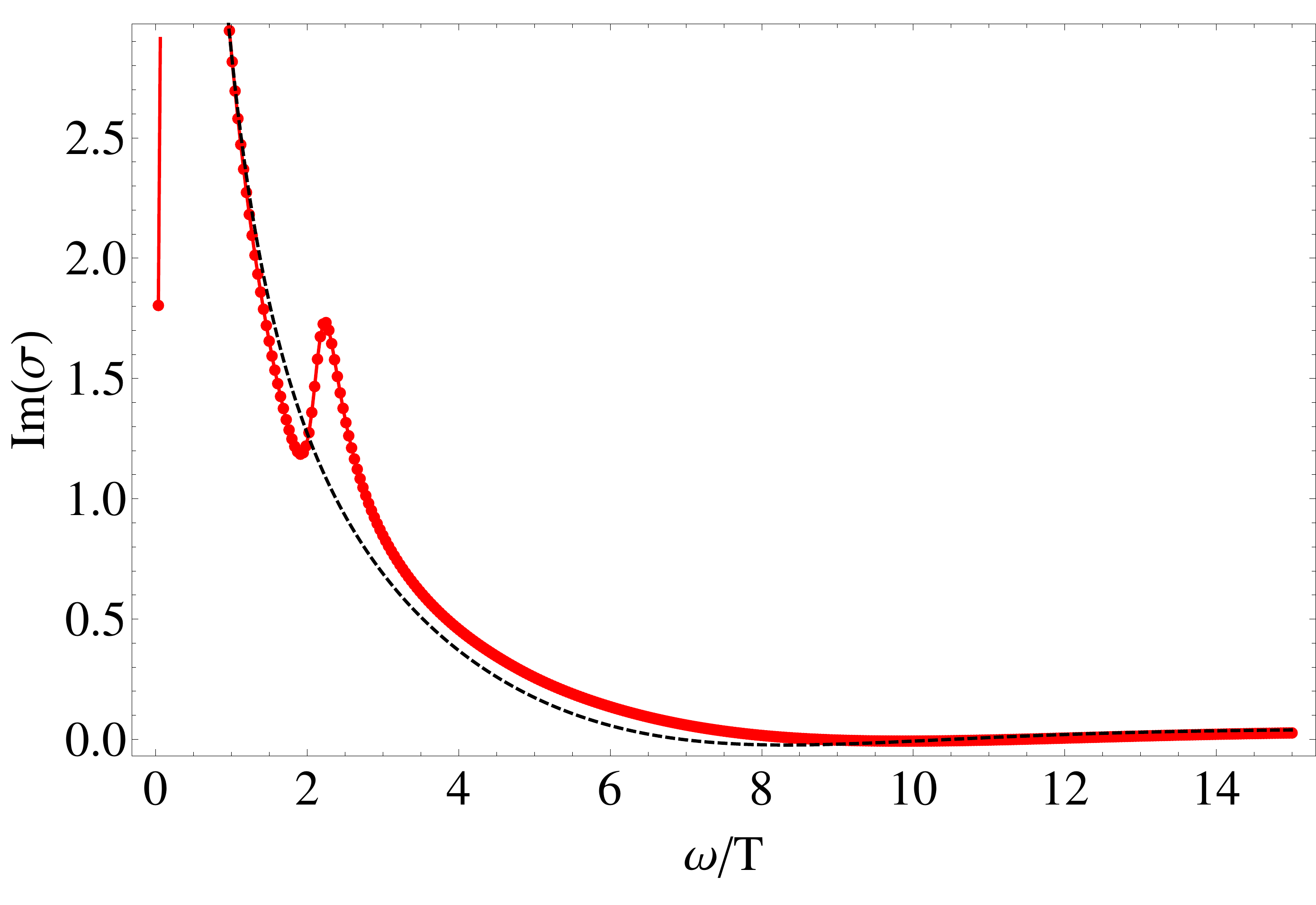}\\ \hspace{0.1cm}
\caption{\label{Resonances1}The resonances of the conductivity for a lattice with $A_0=0.6$, $k_0=1$ and $Q=1$.} }
\end{figure}

\begin{figure}
\center{
\includegraphics[scale=0.28]{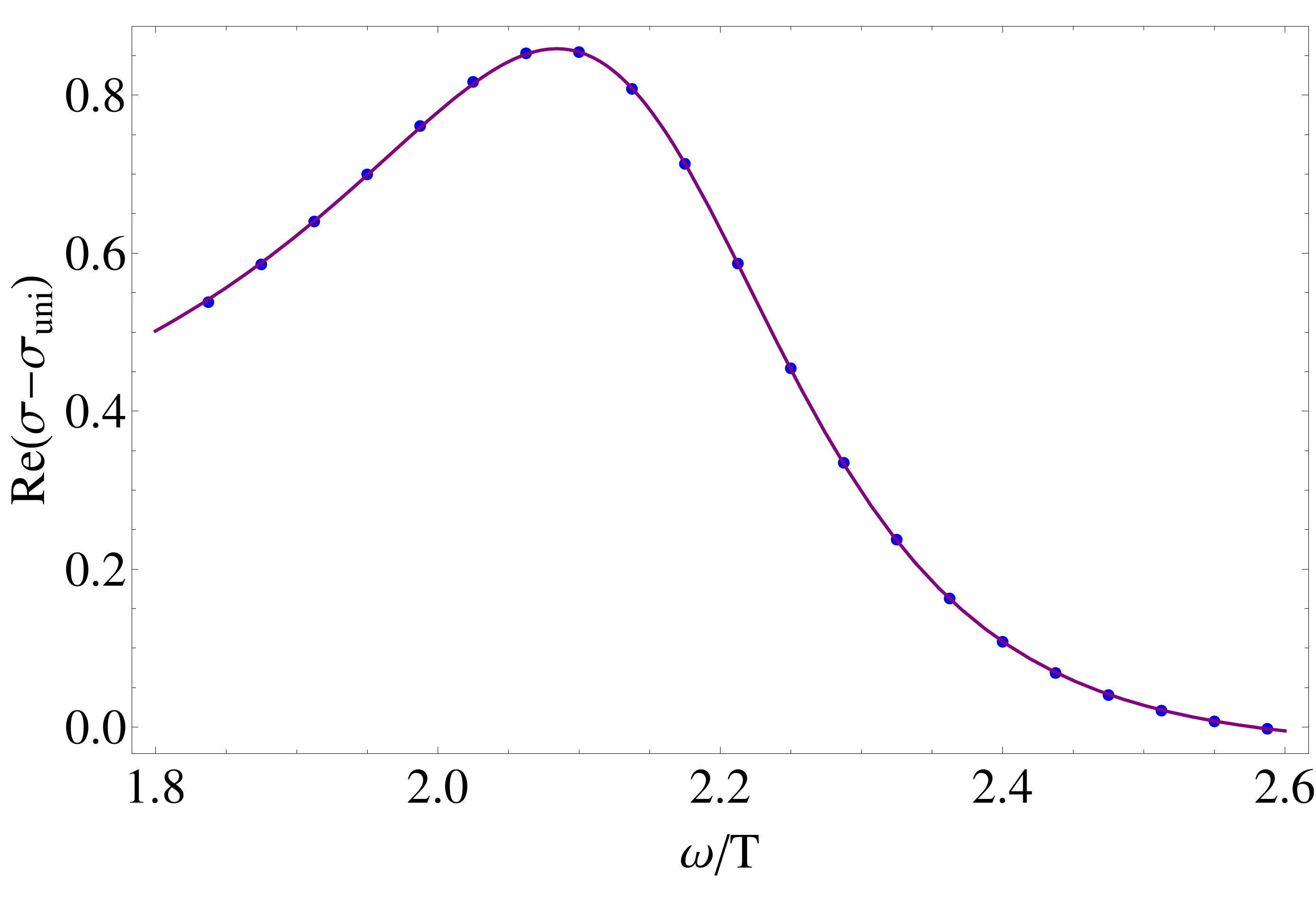}\hspace{0.1cm}
\includegraphics[scale=0.28]{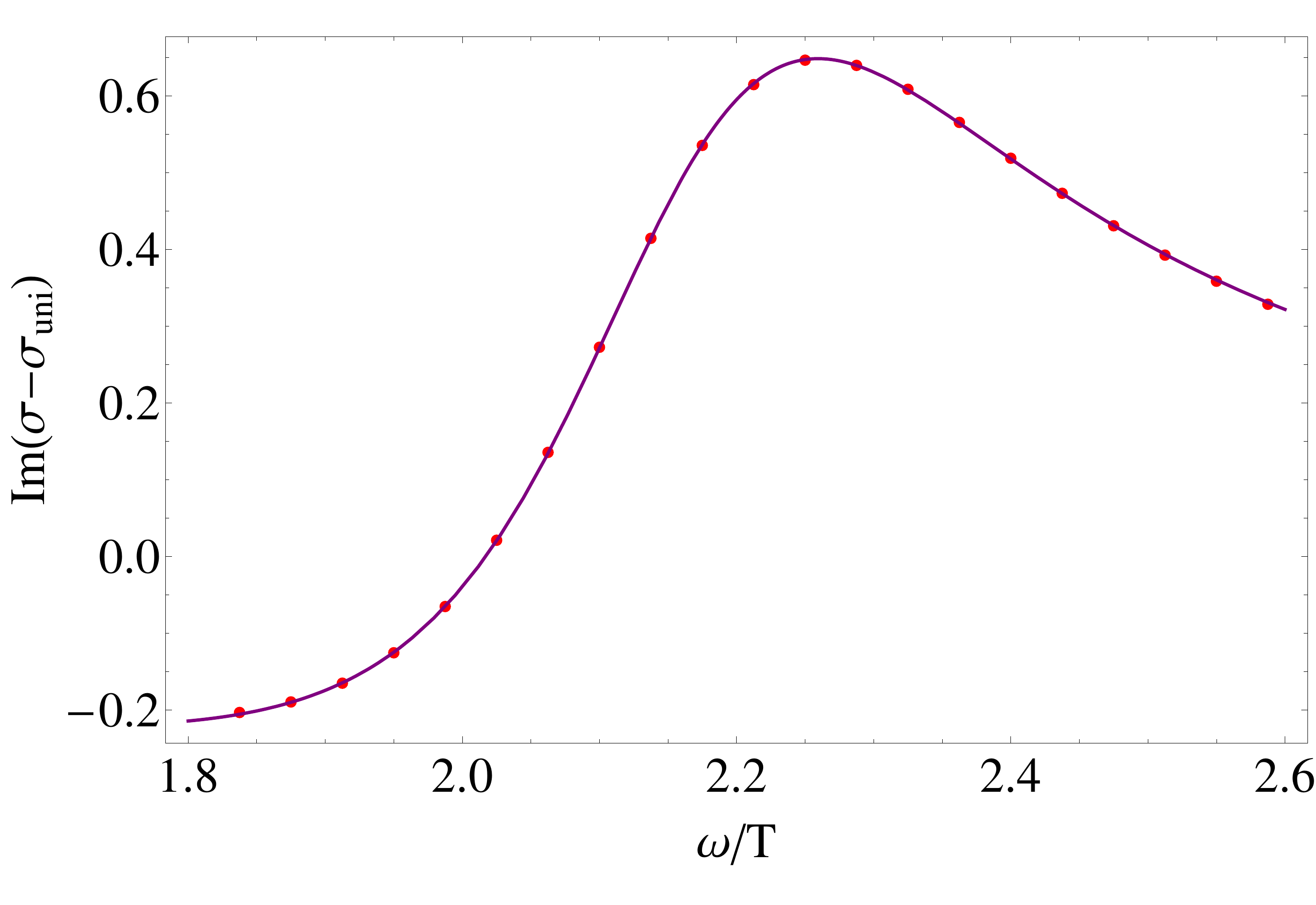}\\ \hspace{0.1cm}
\caption{\label{Resonances2}The resonances of the conductivity with the homogeneous background subtracted off. The data points are well fit by (\ref{resonances}).} }
\end{figure}

When the lattice amplitude is large enough or the temperature is
low enough, we will observe another interesting feature of the
optical conductivity, namely the resonance, which has previously
been observed in the context of holographic conductivity in
\cite{Horowitz:2012gs}\footnote{We thank J. Santos for drawing our
attention to this phenomenon.}. Such a resonance can be attributed
to the excitation of the quasinormal modes of the black hole with
an integer multiples of the lattice wavenumber. In
FIG.~\ref{Resonances1}, we show a clear example with $A_0=0.6$,
$k_0=1$ and $Q=1$. In order to see the detailed structure of the
resonances, in FIG.~\ref{Resonances2} we subtract the homogeneous
background from the conductivity, and we find that the data can be
fit by
\begin{equation}
\sigma(\omega) =\frac{G^R(\omega)}{i\omega}=\frac{1}{i\omega}
\frac{a+b(\omega-\omega_0)}{\omega-\omega_0},\label{resonances}
\end{equation}
where $G_R(\omega)$ is the retarded Green function which has a
pole at the complex frequency $\omega_0$ and $a, b$ are complex
constants. In FIG.~\ref{Resonances2}, our fitting fixes the
quasinormal mode frequency to be $\omega_0/T=2.2-0.21i$.


In the end of this section we intend to present some remarks on
what we have observed in our lattice model and hopefully such
observations are valuable for further investigation on linking the
holographic lattice with the experiments in condensed matter
physics.

\begin{figure}
\center{
\includegraphics[scale=0.37]{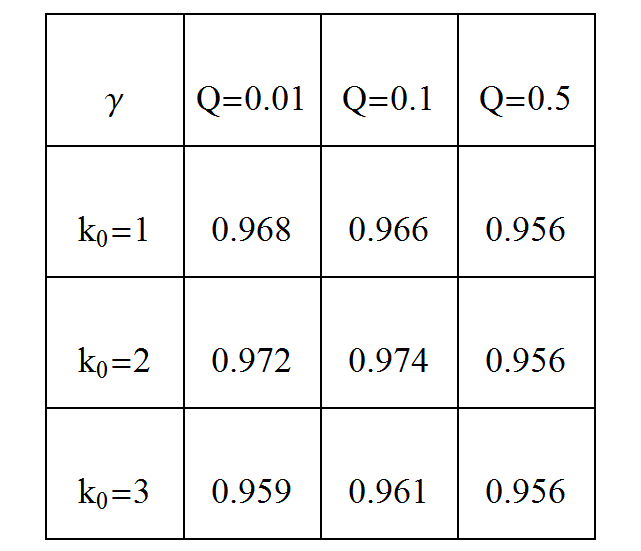}\hspace{0.1cm}
\includegraphics[scale=0.29]{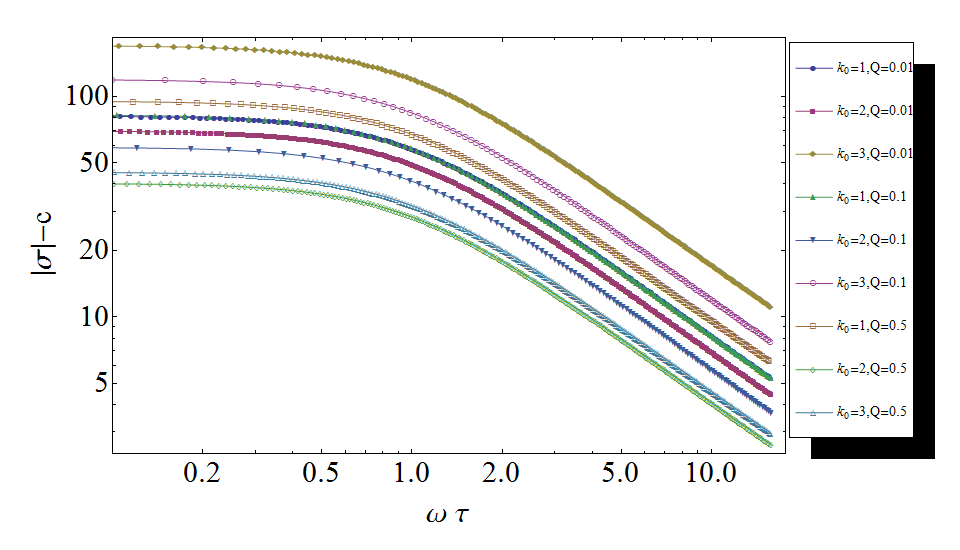}\\ \hspace{0.1cm}
\caption{\label{gamma2}The values of the exponent $\gamma$ with
different $Q$ and $k_0$ when fitting the data in the regime
$5\leq\omega \tau\leq 16$, where we have fixed $k_0/A_0=5$.} }
\end{figure}

\begin{itemize}
    \item First of all, the value of the exponent is universal in the sense that it is
    independent of the parameters in the model. However, it {\it does} depend on what the intermediate regime of the frequency we would pick out.
    In particular, we find it is quite sensitive to the initial point of the intermediate interval. For instance, if we select an interval
    $1.5\leq\omega \tau\leq 8$ as the intermediate regime, we find the the exponent has shifted to $\gamma \simeq 0.71$, irrespective of the
    values of the parameters. Keep going further, for the interval $5\leq\omega \tau\leq 16$ which perhaps is far from the Drude peak, we find the data of modulus
    can still be well fit with a power law formula, while for this regime
    the exponent increases up to $0.96$ in our model (See FIG.\ref{gamma2})\footnote{ The fact that the exponent is close to one probably implies that
    our model contains a regime which could be viewed as a graceful exit from the Drude
    regime since from Eq.(\ref{drude}) we have $|\sigma(\omega)|\sim
    K/\omega $ as $\omega\tau\gg 1$.}.
    Mathematically such a dependence should not be
    quite surprising since we have introduced an offset $C$ such that we have a lot of room to fit the data with a power law
    formula. Physically, our understanding is that such
    holographic scenarios contain a lot of room to fit data in experiments not only for curprate but also for other materials,
    provided that we select an intermediate frequency regime appropriately.

    \item Secondly, the involvement of offset $C$ is not desirable
    from the side of condensed matter experiments. We find in
    general such an offset can not be avoided in order to fit the power law
    of the conductivity. But in some subsets of the parameter
    space, we do find that the data can be fit well even without the
    offset $C$.

    \item Thirdly, the resonance is an independent phenomenon from the power law behavior. Both of them could occur at different frequency regimes.
    While once one increases the lattice amplitude, the position of the resonance may enter the intermediate frequency regime such that the power law behavior
    presented in FIG.\ref{gamma2} may be swamped.
\end{itemize}

\section{Conclusions and discussions}\label{SConclusions}
In this paper, we have constructed a holographic lattice model in
Einstein-Maxwell-Dilaton theory and investigated the lattice
effects on the optical conductivity of the dual field theory on
the boundary. Due to the breaking of the translational symmetry in
the presence of the lattice, the delta function previously
appeared in the real part of the conductivity is smeared out and
the behavior of the conductivity at the low frequency regime can
be exactly described by a simple Drude formula. We have also
observed a power law behavior for the modulus of the conductivity
at an intermediate frequency regime, as that firstly found in
\cite{Horowitz:2012gs,Horowitz:2012ky,Horowitz:2013jaa}. Moreover,
a resonance phenomenon has been observed when the lattice
amplitude is large or the temperature is low enough. Our results
furthermore justify the effects of the violation of translation
symmetry in the presence of the lattice and test the robustness of
the power law behavior at the intermediate regime.

A lot of work can be further done in this direction. First of all,
as we stated the zero temperature limit is out of touch in our
current paper. Secondly, we can construct a holographic
superconductor model with lattice in this theory and study the
appearance of the energy gap in superconducting phase. More
importantly, we expect that we can understand the simple Drude
form (\ref{drude}) in this model through an analytical treatment,
in which we can approximately obtain an analytical solution of the
conductivity at $\omega\rightarrow 0$ by matching the inner region
and outer region as suggested in
\cite{Zaanen:2012Fe,Davison:2013Ma,Gubser:2012Fe,HongLiu2}.
However, it is difficult to have an analytical treatment at the
intermediate frequency regime by employing the matching method
directly since the expansion of the perturbations at intermediate
frequency is probably problematic. Our work on above issues is
under process and we expect to report our results elsewhere in
future.

\begin{acknowledgments}

We specially thank Jorge Santos for kind correspondence and very
valuable suggestions. We are also grateful to Xianhui Ge,
Chaoguang Huang, Yu Tian, David Tong, Xiao-ning Wu and Hongbao
Zhang for helpful discussions and responses. This work is
supported by the Natural Science Foundation of China under Grant
Nos.11275208,11305018 and 11178002. Y.L. also acknowledges the
support from Jiangxi young scientists (JingGang Star) program and
555 talent project of Jiangxi Province. J.W. is also supported by
the National Research Foundation of Korea(NRF) grant funded by the
Korea government(MEST) through the Center for Quantum
Spacetime(CQUeST) of Sogang University with grant number
2005-0049409. He also acknowledges the hospitality at APCTP where
part of this work was done.

\end{acknowledgments}

\end{document}